# Design and Implementation of a Wireless Sensor Network for Smart Homes

Ming Xu[1], Longhua Ma[1], Feng Xia[2], Tengkai Yuan[1], Jixin Qian[1], Meng Shao[3]

[1]Department of Control Science and Engineering, Zhejiang University, Hangzhou 310027, China
e-mail: lhma@iipc.zju.edu.cn
[2]School of Software, Dalian University of Technology, Dalian 116620, China
e-mail: f.xia@ieee.org
[3]Computer Centre, Hangzhou First People's Hospital, Hangzhou 310006, China

*Abstract*—**Wireless sensor networks (WSNs) have become indispensable to the realization of smart homes. The objective of this paper is to develop such a WSN that can be used to construct smart home systems. The focus is on the design and implementation of the wireless sensor node and the coordinator based on ZigBee technology. A monitoring system is built by taking advantage of the GPRS network. To support multi-hop communications, an improved routing algorithm based on the Dijkstra algorithm is presented. Preliminary simulations have been conducted to evaluate the performance of the algorithm.**

## I. INTRODUCTION

Wireless technologies have been developing rapidly in these years. The obvious advantage of wireless transmission is a significant reduction and simplification in wiring and harness [1]. Many communication technologies, such as IrDA, Bluetooth and ZigBee, GSM/GPRS (General Packet Radio Service), etc., have been developed for different situations. Nowadays, a kind of real time systems in which multiple sensors connected simultaneously to one gateway unit become necessary, and they are transformed into wireless sensor networks (WSNs).

In previous work, much research has been done using wireless sensor technologies. Literature search indicates that applications using wireless sensor technologies have already existed in the following four fields [2]:

(1) Home automation and remote monitoring of houses. For example, Liang *et al* [3] developed a system of wireless smart home sensor network based on ZigBee and PSTN (Public Switched Telephone Network) technologies.

(2) Environmental monitoring, including humidity, temperature and radiation. For instance, Rosiek and Batlles [4] presented a system of data-acquisition from remote meteorological stations using the mobile communication networks (more specifically, GPRS).

(3) Fault tracking and fault management. For example, in [5], the authors developed an online diagnosis and real time warning system for vehicles using 3G technologies and GPRS communications.

(4) Health monitoring. For instance, Monton *et al* [6] designed an e-health approach to monitoring data of specific population, such as electroencephalograms, electrocardiograms, electromyograms and so on, which uses ZigBee-based WSNs. ZigBee is particularly suited for the implementation of a wide range of low cost, low power consumption, reliable control and real-time monitoring applications within the smart home situations. The above-mentioned four application areas are also closely related to the design of a WSN for smart homes.

In the past, research in smart home and in-home applications was often limited to ZigBee technology, and gradually other long-distance network technologies such as PSTN [3] and GSM [7] are adopted. It turns out that the use of these technologies makes information more accessible. This would significantly improve people's living quality. However, as a traditional wired network, PSTN has some problems, such as unsatisfactory security assurance, inconvenience and high cost. Therefore we need a new solution. Among other choices, the GPRS technology can solve these problems. Thanks to its unprecedented ubiquity, GPRS is now available almost anytime and anywhere, for anybody (being served). Furthermore, the GPRS network has a highly secure infrastructure, which makes sure that the information sent or received cannot be stolen [7]. Based on these observations, we propose to develop a new system utilizing ZigBee sensor networks and the GPRS network connecting the ZigBee networks to the application server. Details about the design and implementation of the system are presented in the following sections.

## II. SYSTEM DESIGN

As shown in Fig. 1, a star-mesh hybrid topology is usually used in the smart home system. It mainly contains the following components:

(1) ZigBee network coordinator: This special node takes the responsibilities of controlling data communications, establishing communication links and protecting equipments inside the network.

(2) ZigBee node: A node is mainly composed of various sensors and a ZigBee wireless module. In practice, nodes can be deployed to establish a network with a star-like, mesh, or hybrid topological structure. In the monitoring area, ZigBee nodes are scattered according to the distance and all sensor data can be sent to the network coordinator though the network.

(3) GPRS network: Data generated within the ZigBee network is transferred to the monitoring center via the GPRS network and the Internet.

(4) Monitoring center: A computer/server in the monitoring center is used to manage the data generated by all Zigbee networks.

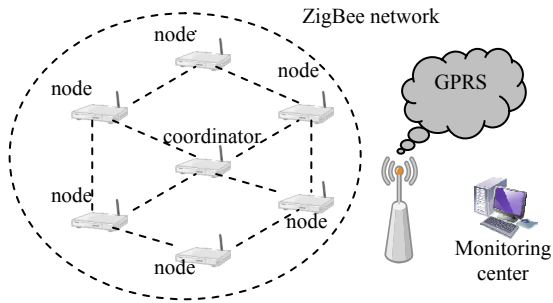

Figure 1. Network topology of smart home systems

In our daily applications, several ZigBee networks would overlap each other and the nodes of the network may breakdown [8, 9]. Therefore characteristic identification of the node should be considered in the design to distinguish the networks. Redundancy nodes are needed to deploy and the mesh topological structure should be optimized. What's more, the routing-algorithm should be examined to ensure the communications among nodes.

*A. Hardware Design*

The network node and the coordinator are key components of the system. In this work, we assume that no matter where the user is, the coordinator will always be connected to the monitoring center/server via a computer that can access the Internet or the GPRS network. It can obtain all messages exchanged between the server and the network. When the server sends out a command, the CPU of the network coordinator will read the content of the command and get the details by analyzing it, such as turning on the air conditioner or refrigeration. The main control program within the network coordinator writes the details to the ZigBee module through serial ports. Then the ZigBee module will be responsible for sending the messages to the family network.

From Fig. 1 we can see that development of the network coordinator and the ZigBee node is the most important task for hardware system design. These two components are basically identical. The only difference is that the latter has the function of GPRS communications while the former does not. Therefore, we will focus on describing the design of the ZigBee network coordinator in this paper. According to the above description, we can find that the microcontroller (MCU), the ZigBee module, and the GPRS module are the most important parts of the network coordinator. The hardware construction of our coordinator node adopts MSP430F149 as MCU (from TI), ETRX2 ZigBee module, and MC52i GPRS module (from Siemens).

The three modules can connect and communicate with each other through serial ports.

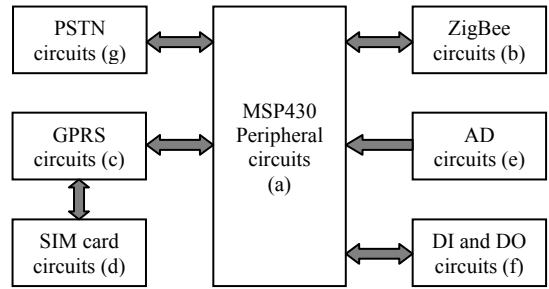

Figure 2. Internal structure of coordinator

As shown in Fig. 2, parts (a), (b) and (e) constitute a typical ZigBee node. Adding parts (c) and (d) to the node, it becomes a ZigBee coordinator. We add parts (f) and part (g) for additional functions. Part (f) can control some target with 2A current relay and monitor the switch mode of the target. Part (g) can generate and decode PSTN signals. Home secure alarm mainframe can be connected to the tip and ring port. When the mainframe alarms, part (g) can extract alarm information of Contract ID (CID, one popular alarm protocol) code quickly, and the coordinator will send an alarm message to the monitoring center.

The major function of the coordinator in remote communications is as follows. The GPRS communication unit connects to the MCU through the RS232 connector, and is responsible for data transmission between the node and the monitoring server. The data sent by the server will get into the GPRS communication module by antenna. Useful data will be obtained through analyzing the TCP/IP agreement. Response data of the MCU will be modulated to GSM signal by the GPRS module and be sent to the server via the internet using the TCP/IP protocol.

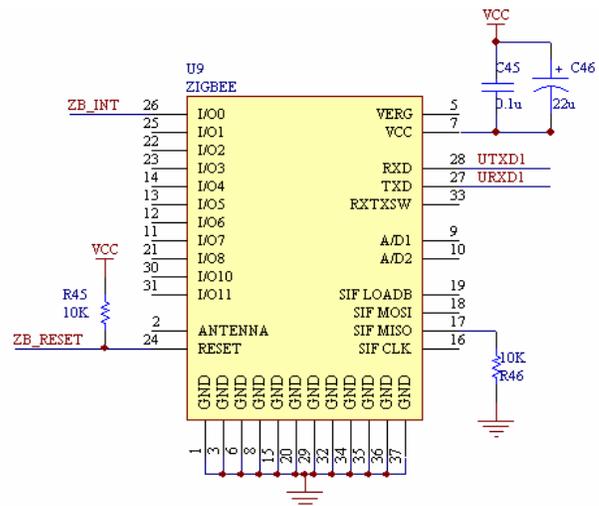

Figure 3. ZigBee circuit diagram

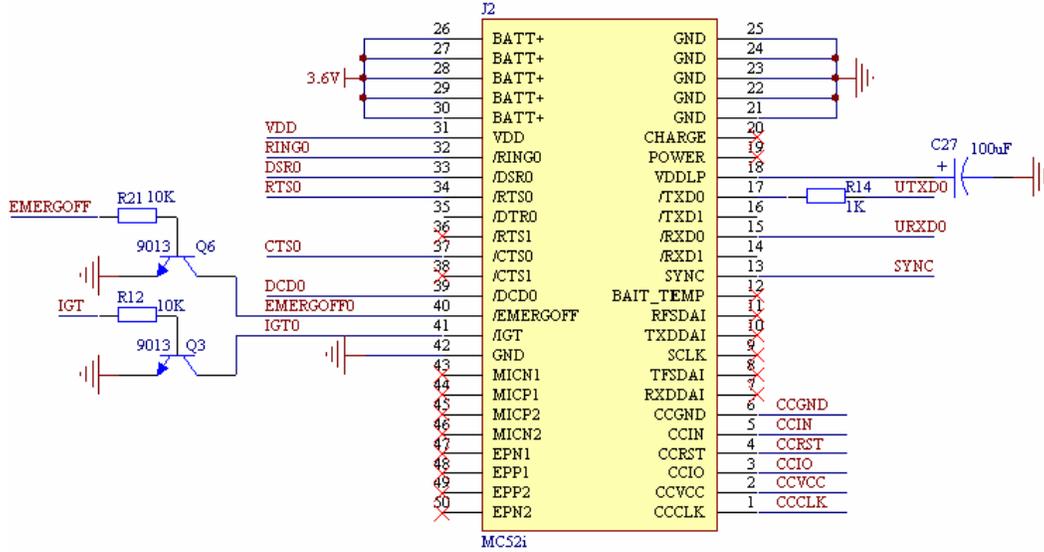

Figure 4. GPRS circuit diagram

We have designed peripheral circuits according to the functional requirements, and developed a network coordinator by integrating the ZigBee coordinator node and the GPRS module together on a PCB board. Given in Fig. 3 and Fig. 4 are some of the circuit diagrams designed for the system. Fig. 5 shows the hardware board of the coordinator node we developed.

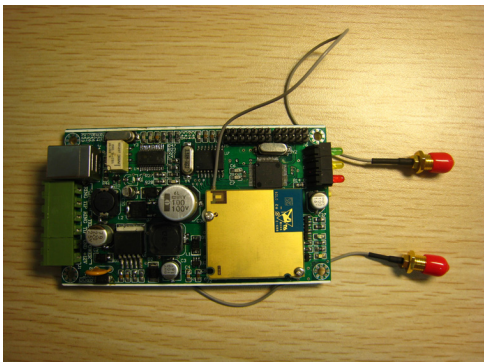

Figure 5. Coordinator node developed.

## B. Software Design

### 1) Monitoring Software

The monitoring software on host station (i.e. monitoring server) adopts C/S architecture based on Socket communication mechanisms of TCP/IP protocol. It is written by C# language, using ACCESS database.

Built on .Net software platform, this software features independence of platforms and excellent expandability. The whole monitoring system of the host station mainly consists of two parts: database management server and system management server. The database management server includes database server and databases. The database management server serves to manage the interaction between various modules and databases, to establish databases, and to connect to other modules. The system management server interacts with the users. The system management server provides a human-machine interface, through which users can configure system parameters, monitor real-time data, inquire about historical records, etc. The framework of the software is illustrated in Fig. 6.

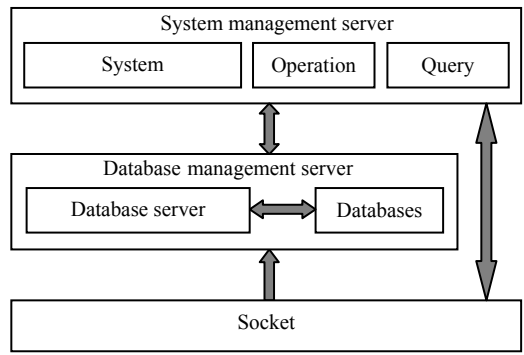

Figure 6. Structure of host station software

### 2) Node Software

The network (coordinator) node software realizes the collection and transmission of data. Fig. 7 shows the block diagram of the software. Fig. 7(a) is the main procedure flowchart and Fig. 7(b) is the interruption procedure flowchart. As for sensor nodes, the program realizes functions such as data sampling, A/D calculation, I/O control, timed sending, and timed hibernating. As for router nodes and coordinator nodes, it mainly realizes the function of data forwarding and path routing. The router nodes and the coordinator nodes have the capacity of collecting sensor data. Therefore they can also be treated as sensor nodes.

During the interaction between coordinator nodes and the GPRS network, both of them should follow the same datagram protocol in order to enable the host station to analyze the message more easily.

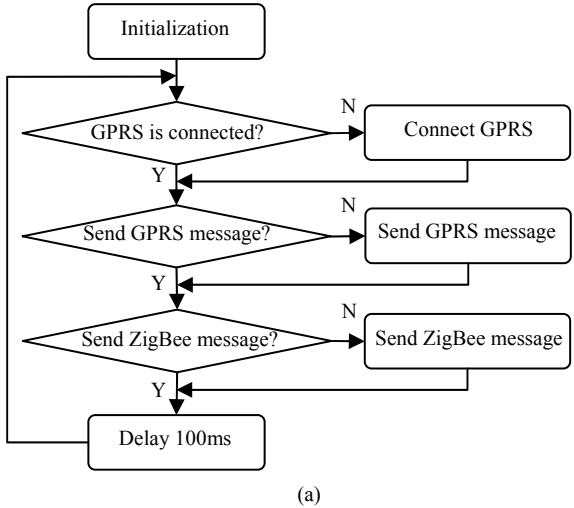

(a)

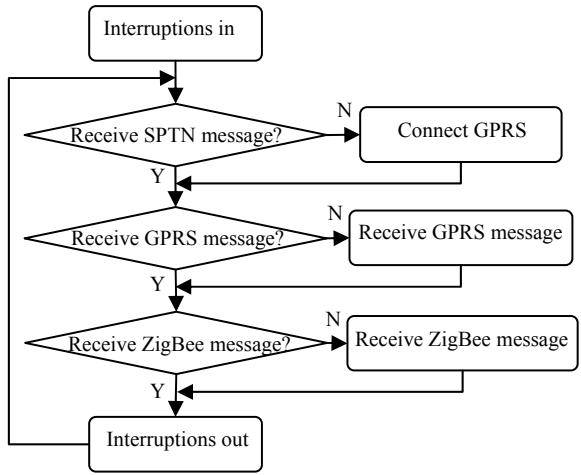

(b)

Figure 7. Flowchart of node software: (a) main procedure; (b) interruption procedure.

## III. ROUTING ALGORITHM

The smart home system using the WSN developed above can be modeled as a (wireless) network, and the routing point is the node in the network. The traditional Dijkstra algorithm [10] generates the shortest path according to the order of increasing path length, and greedily searches path based on the edges connected with nodes. However, there is no edge in the wireless network. Therefore we propose an improved Dijkstra algorithm for the WSN, which obtains the shortest path in the network.

Assume that there are $n$ nodes in a wireless network, and the location of each node is available. Then we can get the table of distances between nodes by using the following algorithm [11]:

(1) A node is arbitrarily selected as the root node. After initialized, it will send messages to surrounded nodes asking for their IDs and location information.

(2) In response to Step (1), the remaining $n$-1 nodes send their IDs and location information to the root node.

(3) The distance table is created after the root node has received all the information of remaining $n$-1 nodes.

As an example, Fig. 8 gives a simple wireless network. The corresponding distance table is shown in Table I.

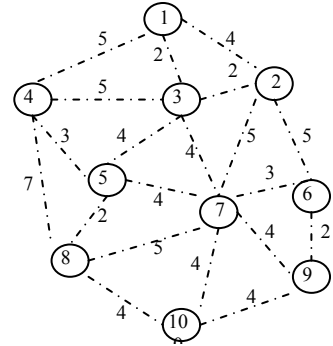

Figure 8. An example network of 10 nodes

TABLE I. DISTANCES BETWEEN NODES

| Node ID | 1 | 2 | 3 | 4 | 5 | 6 | 7 | 8 | 9 | 10 |
|---|---|---|---|---|---|---|---|---|---|---|
| 1 | 0 | 4 | 2 | 5 | 5 | 7 | 6 | 8 | 9 | 9 |
| 2 | 4 | 0 | 2 | 7 | 6 | 5 | 5 | 7 | 7 | 8 |
| 3 | 2 | 2 | 0 | 5 | 4 | 5 | 4 | 6 | 7 | 8 |
| 4 | 5 | 7 | 5 | 0 | 3 | 8 | 7 | 7 | 9 | 8 |
| 5 | 5 | 6 | 4 | 3 | 0 | 6 | 5 | 2 | 7 | 5 |
| 6 | 7 | 5 | 5 | 8 | 6 | 0 | 3 | 8 | 2 | 5 |
| 7 | 6 | 5 | 4 | 7 | 4 | 3 | 0 | 5 | 4 | 4 |
| 8 | 8 | 7 | 6 | 7 | 2 | 8 | 5 | 0 | 6 | 4 |
| 9 | 9 | 7 | 7 | 9 | 7 | 2 | 4 | 6 | 0 | 4 |
| 10 | 9 | 8 | 8 | 8 | 5 | 5 | 4 | 4 | 4 | 0 |

After the table of distances is created, we use an improved Dijkstra algorithm to deduce the optimal path. Some symbols and notations used in the algorithm are listed in Table II.

TABLE II. SYMBOL DEFINITION

| Symbol | Description |
|---|---|
| $n$ | Number of network nodes |
| $v$ | Sending node |
| $w$ | Receiving node |
| $k$ | Transmission radius |
| $u$ | Relay node |
| $s[i]$ | Visit mark (if not visited, set $s[i]$=0; else set $s[i]$=1) |
| $cost[i][j]$ | Distance between node $i$ and node $j$ |
| $dist$ | Distance corresponding to the optimal path from $v$ to $w$ |

The improved Dijkstra algorithm is described in the following:

(1) Initialization: *num*=0, *dist*=+∞, *s*[*i*]=0, (*i*=0,1,…*n*);

(2) If *cost*[*v*][*i*] ≤*k*, then set *s*[*i*]=1, (*i*=0,1,…*n*);

(3) If *s*[*w*]=1, then *dist*=*cost*[*v*][*w*]; otherwise go to step (4);

(4) ∀ *i* ∈ {*s*[*i*]=1}, *u*[*num*]=1, set *s*[*j*]=1 when *j* ∈ {*cost*[*u*[*num*]][*j*] ≤*k*}, (*j*=0,1,…*n*);

(5) If *s*[*w*] ≠ 1, then *num*++, repeat step (4); otherwise record the path that fulfills *s*[*w*]=1, and

$$dist[num] = \cos t[v][u[0]] + \sum_{j=0}^{num-1} \cos t[u[j]][u[j+1]] + \cos t[u[num]][w];$$

(6) For *num*=*num*+1, repeat steps (4) and (5) until all the paths that fulfill *s*[*w*]=1 are obtained;

(7) Set *dist*=min{*dist*[*i*], *i*=0,1,…,*num*}, and output the corresponding path.

We implemented the above algorithm and conducted simulations of a network as given in Fig. 8. Setting *k*=5, the optimal path from sending node 1 to receiving node 10 is to pass a relay node 5, which is better than the path of node 1 → node 3 → node 7 → node 10, and reduces the number of node hops. Therefore, the improved Dijkstra algorithm can solve the problem of optimal path selection in a wireless network, thus providing a feasible routing solution for smart home systems.

To evaluate the performance of routing algorithms, a general concern would be the energy consumption of the network nodes. If a node is put into sleep as long as it has no data to receive or send (and all state switching overheads are negligible), its energy consumption will be approximately proportional to the number of times it is visited (i.e. the number of packets it has transferred). In this case, it is possible to examine the relative energy consumption of all nodes by observing the number of times each node is visited.

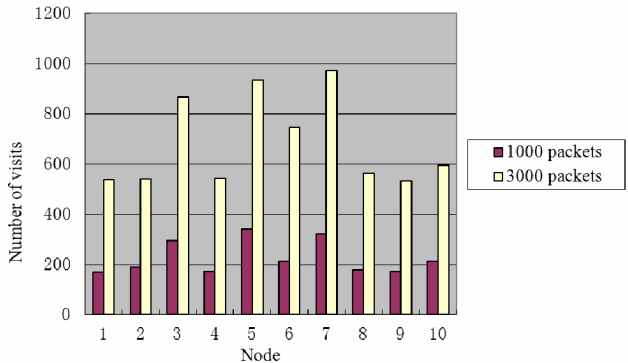

Figure 9. Number of times each node is visited.

In simulations the sending node and the receiving node are randomly generated. Fig. 9 gives the results with data transformation of 1000 and 3000 times. Noticing that nodes 3, 5 and 7 have more visited times, we can conclude that nodes 3, 5 and 7 consume more energy. This observation could be helpful when placing nodes in a smart home.

IV. CONCLUSIONS

This paper focused on development of the wireless sensor node and the coordinator for smart home systems based on ZigBee technology. Both hardware design and software design have been described in detail. A monitoring system is also built using the GPRS network. To address the problem of routing for multi-hop communications in smart home wireless sensor networks, an improved algorithm based on the Dijkstra algorithm has been presented. The performance of the algorithm has been evaluated through preliminary simulations. Our future work is to test and apply the whole system in practice.

ACKNOWLEDGMENT

This work is supported in part by Natural Science Foundation of China under Grants No. 60474064 and No. 60903153, Zhejiang Provincial Natural Science Foundation of China under Grant No.Y107476 and No.R1090052 and the Fundamental Research Funds for the Central Universities.